# Teaching Information Security Management in Postgraduate Tertiary Education: The Case of Horizon Automotive Industries

**Teaching Case**

**Atif Ahmad**
School of Computing and Information Systems
The University of Melbourne
Parkville, Victoria, Australia
Email: atif@unimelb.edu.au

**Sean B. Maynard**
School of Computing and Information Systems
The University of Melbourne
Parkville, Victoria, Australia
Email: seanbm@unimelb.edu.au

**Sameen Motahhir**
Center for Digital Humanities, Department of Languages and Linguistics
Information Technology University
Lahore, Pakistan
Email: sameen.motahhir@itu.edu.pk

## Abstract

Teaching cases based on stories about real organizations are a powerful means of storytelling. These cases closely parallel real-world situations and can deliver on pedagogical objectives as writers can use their creative license to craft a storyline that better focuses on the specific principles, concepts, and challenges they want to address in their teaching. The method instigates critical discussion, draws out relevant experiences from students, encourages questioning of accepted practices, and creates dialogue between theory and practice We present 'Horizon', a case study of a firm that suffers a catastrophic incident of Intellectual Property (IP) theft. The case study was developed to teach information security management (ISM) principles in key areas such as strategy, risk, policy and training to postgraduate Information Systems and Information Technology students at the University of Melbourne, Australia.

# 1　Act 1

**Present time, somewhere in the north of Europe**

Wednesday the 1st of March started like any other day. Anders Svensson, CEO of Horizon Automotive, parked the firm's latest model hypercar out the front of a tall non-descript building. He hurried up the steps, skipped through the revolving door and made his way to a corner office on the top floor. "Good morning Erik," he smiled at a middle-aged bespectacled gentleman. Erik had his head down close to the keyboard – he was typing cautiously with his two index fingers. He would periodically look up to squint at his large screen.

Awkward silence. … Erik looked up towards Anders. "Hello Anders, I have some unfortunate news." "What news?" Anders stared at Erik. He could feel a knot forming in his gut. "I'm afraid we won't be placing that order for your NextGen model this year." Erik looked half apologetic. "Why? If it's a matter of price we can negotiate a volume discount if you increase your order to at least 10 units?" "No, no… it's not a problem of price," Erik took a deep breath and leaned back on his expensive leather chair. "Then?" Anders looked perplexed. "We've decided to place the order with a different firm." Erik braced for Anders' reaction.

Anders jaw almost dropped to the floor. "What do you mean you've awarded the contract to someone else? We are the leading manufacturer of performance cars in the country. We're the ONLY manufacturer of performance cars in the country." "Yes… Well… that's true," stuttered Erik. "But we've found another firm that can fulfill our needs. And at a price you won't be able to match." Erik could be so matter of fact if he wanted to. Anders world was spinning. Horizon dominated the super car market in this part of the world. The firm only manufactured 20 to 25 cars every year, but each sold for well over 1.5 million dollars – US dollars that is.

"You've found a firm that produces engines that weigh around 200kg but can produce over 900 hp and can accelerate a car from 0 to 100 km/h in 2.5 seconds flat?" "They say they can," blinked Erik. "And they are almost as pretty as yours. In fact, come to think of it, there's even a faint resemblance to your cars. It's as if they are siblings or cousins perhaps." "Who?" uttered Anders, meekly. "Who what?" muttered Erik. "What is the name of the firm?" Anders could feel the heat rising to his face. "I'm sorry. I can't say. But it's not a local firm. In fact, it's not even a European firm." "What's the name," Anders insisted. Erik folded his arms and stared into space. Anders could almost feel his face becoming bright red. "Give me the name!" Erik got up, peered down the corridors and then shut the door. A few seconds later Anders stormed out of the office - his chest heaving, his anger palpable. He pulled out his cell phone and hit the first speed dial button. "Patrick, I need your help. I need you to find what you can on a firm named FC Design."

# 2　Act 2

**Present Time, Horizon Automotive's offices in Sweden**

Anders sat in his posh penthouse office with Lena, his trusted confidante, and VP Research & Development. A manila folder lay on the mahogany table in front of them. It was marked 'From Patrick Chau - Nordic Risk Intelligence'. Lena carefully undid the clasp and lifted a series of high-resolution pictures. "Oh, dear God," Anders was staring at the spitting image of the NextGen Horizon super car. The car was still under development and no living soul had seen the design outside of the R&D team, and him of course. "How could it be?" Lena looked up at Anders. "What did you say the name was?" "FC Design". "Wasn't that the firm that hired one of the engineers from our R&D team about a year ago? Anders stared at Lena for a moment, his jaw tightened. He picked up the phone to his IT Manager. "Kurt – I need you to review the logs of all R&D systems from about a year ago. I want to know if Emil accessed any files from the NextGen project before he left."

# 3　Act 3

**A few years prior to the present time, Dubai (United Arab Emirates)**

Oliver was quite at home strolling barefoot on the white sands of Dubai's Jumeirah beach. The sun was penetrating, as usual, but the cool breeze made it somewhat tolerable. "Oliver?" a voice came from behind him. A middle-aged gentleman dressed casually but fashionably in a sports blazer, a t-shirt, shorts, and expensive sunglasses cautiously extended his hand. Oliver shook the hand and pointed to a nearby cafe.

Oliver took the first table under the shade of a large yellow umbrella, thankful to be out of the sun. The client sat opposite him. He took off his glasses and laid them on the table in a slow deliberate motion, as if he were reluctant to reveal his face. The almond-shaped eyes focused laser-like at Oliver. The client was clearly Asian, extremely wealthy by the way he dressed, and appeared to be very educated from the way he constructed his sentences. That's about all Oliver could tell at the moment. The client reached into his front pocket and pulled out a business card, carefully laying it on the table in front of Oliver. The card read "FC Design Automotive Industries". On the second line it said "Research & Development" in bold lettering. No name specified on the card. Oliver glanced at the contact details and address at the bottom of the card -committing the details to memory.

"Call me John," the client showed no emotion. "How can I help?" Oliver was curious. "We would like to procure your services," the client had a distinctive Canadian accent. "To do what exactly?" Now Oliver was intrigued. The client's hand went into the inside pocket of his jacket pulling out a photograph the size of a greeting card. "Do you recognize this car?" Oliver raised his eyebrows. "Yes, I do. It's an Horizon Altitude. Top of the line performance vehicle." "Excellent," said the client returning the photograph to his pocket.

"The incoming CEO of FC Design has outlined an ambitious new vision for the company. He would like the firm to compete in the hypercar market in ten years. Recently, engineers from our R&D division visited the manufacturing division of Horizon. We were very impressed with their… capability." Oliver remained silent. The client continued… "We have the financial resources, we have the factory space, and a highly skilled engineering workforce with considerable experience in building passenger vehicles. But engineering a hypercar requires a higher order capability. To start with, we need you to collect as much information as possible about their next generation project - the designs, the specifications of the parts, the names of their suppliers, the pricing of the raw materials, everything. "

"Why the pricing of the raw materials?" "To improve our negotiating position with the suppliers," smiled the client. "Of course," said Oliver sheepishly. "And the matter of my remuneration?" The hand went into the suit jacket once more. Out came an envelope. Inside was a piece of white paper with a figure written on it. Oliver counted the zeros after the leading digit. "Excellent," he smiled.

## 4  Act 4

**Present time, Horizon Automotive's offices in Sweden**

"I've got bad news, unfortunately." Kurt had that scowl on his face again. "Emil copied hundreds of files from the NextGen project server a few weeks before he left the firm." Anders was livid. "How come this is the first time I'm hearing of this?" "Umm… actually you were briefed at the time, but we were distracted by a procurement issue." Anders didn't say anything. He vaguely recalled that Horizon's supply chain was under threat because a key supplier had declared bankruptcy.

"Well - that's an Information Security breach. This is your fault!" Kurt pursed his lips. "Actually - it's not our area of responsibility. We just maintain the availability of IT systems and networks. We have no idea what information is stored on these systems and what is sensitive. There's no 'need-to-know' policy at this firm. Besides, Emil was part of the R&D team, so he had the authority to access these files. "But you are in charge of INFORMATION SECURITY". Now Anders was going red in the face. "Umm… we're responsible for IT security". Retorted Kurt.

"That's the same thing!" roared Anders. "Not exactly." "Damn!" Anders was clearly not happy. "So, who did Emil speak to in the firm, are these people still here? And did they have access to the paper archives? What did they take from there?" "No idea. That's outside the scope of our role. We don't know what's in those archives and how many copies of confidential information might be floating around the organization. We obviously don't know who they're speaking to either." Kurt was becoming a bit uncomfortable.

"So, who's in charge of securing paper files and phone conversations etc.?" Anders was like a dog with a bone. "Ummm… everybody." Kurt blinked. "You mean nobody?" thought Anders to himself. In his mind, he was already calculating the cost of this leak.

## 5  Act 5

**Present time, Horizon Automotive's offices in Sweden**

Patrick Chau took a deep breath, adjusted his horn-rimmed glasses and began his briefing, "FC Design started only ten years ago. The firm produces a range of passenger vehicles for the Asian market. Sales are around 3 million vehicles per annum, so they aren't a smaller player by any means, but they have no market share in Europe.

They clearly don't have the 'know-how' or the technology to build cars with the kind of propulsion, weight-efficient construction and aerodynamics of hypercars...." Anders raised his hand for Patrick to stop. "How does a garden-variety automotive manufacturer graduate to building hypercars in less than ten years?" "Well that's the billion-dollar question, isn't it?" Patrick shrugged. Anders flipped open his laptop and began to scroll through a folder marked 'stolen R&D files'.

After a few minutes of scrolling and thinking Anders leaned back on his chair. He looked at squarely at Patrick, "There's got to be more to this story. The blueprints are not enough. FC Design had to build a supply chain similar to ours. They had to acquire the carbon-fiber to build the car from that supply chain at lower prices than us. Supplier names, pricing, materials identification is not stored on the NextGen servers. These are stored separately on the financial network. So, Emil could not have gotten his hands on that information."

Anders looked quizzically at Lena and Kurt. Lena was busy poring over her personal ledger - staring at her scribbles - paying no attention to Anders or Patrick. Kurt was still scowling...but there was something else. He seemed to have lost the color in his face. "What?" Anders sensed things were about to go from bad to worse. "Remember how a number of Swedish firms, including our IT systems, were hit by a zero-day attack in July 2 years ago?" Anders had already covered his face with his hands. Lena got up and walked out in a hurry, phone in hand.

"Our incident response team reported that the attackers had spent most of their time in our financial network. I'd have to go through their report again and cross-reference the IP addresses, timestamps etc. with the logs but it's quite possible they could have copied a large number of spread sheets, invoices, memos etc." "I'm afraid it gets worse, said Lena stepping back into the room. "I was just speaking to Lars, the head of our physical security team. Apparently, a couple of years ago we had a visit from a delegation consisting of several firms from the other side of the world. FC Design was listed as one of those firms. They were touring Swedish heavy industries. Our engineers reported that some of the delegates, quite possibly from FC Design, had very specific questions about our triplex suspension technology, propulsion systems etc. Some of them were taking pictures with their mobile phones. We eventually confiscated them, but we had to give them back." The dots were beginning to connect in Anders mind.

# 6  Act 6

**Recent Past, London (United Kingdom)**

"Jordan!" shouted Oliver, waving the tall lanky kid over to the park bench where he sat, newspaper in hand. "What did you think of the assignment I gave you?" Jordan shrugged. "Easy mate, E-a-s-y. It's a civilian company, no sweat." "Yes, yes I know," said Oliver impatiently. But I want you to follow standard protocol anyway. What's the plan? And explain it to me in layperson's terms. I don't speak machine language." Jordan scratched his bald head. Then adjusted his glasses on the bridge of his nose. "Okay. I find vulnerabilities on their systems. I need to study the firewalls, servers - see if they have installed all the latest patches. If there aren't installing patches, then it's easy to find a bug to exploit. "Won't that raise an alarm?" Jordan looked at him strangely. "Do you have any idea how many attacks these firms have every day? Every minute? Everyone is door-knocking mate so you can easily hide among them. They won't be paying any attention."

"And if there is no obvious way in then?" Oliver was not convinced. Jordan looked pretty confident, "well... then we hijack some poor executive's laptop. Someone inside the company will use a VPN to remote login, or create a Wifi Access point inside the firm or maybe someone made a mistake wiring the hub so traffic might be going around the firewalls rather than through – or maybe we can go through the photocopier, there are heaps of ways. "Don't you have to be physically present to break in using the wireless hotspot?" Oliver was dotting the i's and crossing the t's. "If it comes to that, then yes – I'll have to fly there."

"And if that doesn't work?" Jordan was getting bored, "worst case scenario - we buy a zero-day vulnerability from the market and design a special malware weapon. We break the malware into small pieces so we can smuggle in the pieces separately through the network perimeter and then reassemble them inside to open a backdoor so we can come in." "How much will that cost?" queried Oliver? Jordan shrugged. "Between $5,000 to $250,000."

Oliver let out a long whistle. "Don't they have an alarm system of some sort to detect if someone has come in?" "You mean an Intrusion Detection System? If we write a special weapon the IDS won't 'trip'". "Why is that?" "IDS' match incoming traffic with signatures in a database – since our malware is unique it will create a unique signature that won't be in any database. We just need to be careful not to borrow code from other malware. "Okay." said Oliver - impressed but trying hard not to show it. "Then what?" "Infiltrate. We make backdoors everywhere so even if they find some, we can come through other entry points. Then we take the information - because we can." Jordan smiled. He rarely smiled. He loved it when a plan came together. "Excellent." Oliver felt satisfied.

## 7　Act 7

**Present time, Horizon Automotive's offices in Sweden**

Anders enters a packed conference room. Seated around the conference table were Kurt (Director, IT Services), Gustav (HR Manager), Lena (VP R&D), Linus (VP, Manufacturing), and Lisa (Corporate Solicitor). Anders looked up at the solemn faces. His cheeks had sunken in and his eyes were weary - he didn't function well without sleep. "This meeting is to discuss the systemic flaws in the way we approach security so we can diagnose the problem and think about what resources we need to marshal. So, for those of you who are not up to speed with the facts of this recent and catastrophic incident, let me recap.

Over the last five to seven years Horizon experienced several security incidents. In hindsight, we now know that these incidents can be linked back to a firm named FC Design. This firm is a mid-range automotive manufacturer without any background in building hypercars. We believe FC Design has very deliberately stolen our IP to erode our competitive advantage and shift the competitive landscape in their favor. They were successful. As a result, in the short-term Horizon has lost a number of contracts, we've got to build a new design from the ground-up, and we're not going to be producing a new car for the next few years at least. The immediate financial hit is in the hundreds of millions of dollars. The bigger problem is that FC Design is going to be a long-term competitor. I'm sure you know what that means for our sales volume, our market share and our profit margin. So... the objective of this meeting is to help us collectively understand how we leaked our IP and what needs to be done to ensure it doesn't happen again. Let me begin by asking a simple question. What's our strategy regarding leakage mitigation?"

Silence. Lisa was the first to speak. Well… we don't have a strategy. And, as the firm's solicitor, I can say we've been taken a number of times to court on the matter. We've leaked confidential contracts by accidentally emailing them to the wrong person, we've had employees downloading God knows how much information on their personal iPads, mobile phones, laptops etc. Must I go on? I've flagged numerous times that we are vulnerable on this front."

Anders began to stroke his chin. "O-k-a-y, not the answer I was looking for but nevertheless thank you. So… let me go back to Kurt. I still don't quite understand why this is NOT an Information Security problem. All the faces in the conference room simultaneously swivelled to focus on Kurt. "Ahem… well… as I said before in a private meeting with Anders and Lena," Kurt spoke in monotone. "All of our computing systems and networks exceed 90% availability. We have firewalls, intrusion detection systems, and anti-virus software that protect us from some pretty sophisticated attacks. We have a master spreadsheet that tracks access privileges - we update that routinely. We also have an incident response team that is ready to mobilize at any time. I could go on, but in general we comply, even exceed industry best-practice standards on IT Security."

Anders was not impressed. His voice was raised - that tended to happen when he got excited. "What good is compliance with best-practice standards?" Anders glowered at Kurt. "Did you identify theft of our entire <bleep> flagship product line as a security issue?" Kurt stared at the floor. He mumbled something. "Your role clearly has 'information' and 'security' in it. So…" Lisa, anticipating the conversation had taken a turn for the worse, raised her hand, "May I jump in for a moment?" Anders took a deep breath. "Please do."

"Well. IT sees their role as maintenance engineers. Their aim is to keep the systems and networks running. That's an important service. We wouldn't be able to run without functioning IT. But they don't see themselves as guardians of our Intellectual Property and trade secrets. So, for example, IT doesn't even know what information is sensitive on the servers. It's all 1s and 0s to them." "So, you are telling me that this is not entirely IT's fault? It's our collective fault? We are all responsible for security? So why haven't we received security training? Gustav?"

"Uh?" Gustav was caught off guard. He wasn't expecting to get involved. "Well we do security awareness training… but…" "…but… the training is only once a year, it covers IT security scenarios - virus attack, system malfunction, etc. nothing like the problem we had here. How do we train employees to report behaviour that we don't know is a security problem until we look back in hindsight?"

Anders was emphatic… "but it was reported! That's the point! Every incident was reported. We just failed to link the incidents together and consider the implications. Why didn't Lena know that Emil had taken those files? Why didn't Kurt know about the sensitive nature of those files? How often do you compare notes about what is going on?"

Anders could see his point resonated with the room. Lena and Kurt looked at each other momentarily. "Not very often. And certainly not about security issues.," Lisa admitted. Anders turned away to peer out through the large window. There was not much to look at actually. Just a massive car park with… cars. "Well, we have a gaping hole in security so we're going to start by appointing a new manager – a Chief Security Officer - this person will have enterprise wide responsibilities to secure our knowledge, information and data regardless of where it is stored and how it is circulated in this firm."

## 8   Act 8

**Recent Past, London (United Kingdom)**

Jordan blinked several times under the gaze of white fluorescent tube lights. "You work in this dump?" queried Oliver as he strode into the massive warehouse - computing equipment strewn across the floor. Jordan ignored the barb. "Over the last 2 days I probed every inch of the Horizon network. I mapped all their systems and noted the version numbers of software, usernames of employees, access level, roles and positions - thank you Facebook and LinkedIn - and I scanned all their drives for keywords. There are a large number of financial documents. I can map out their supply chain in detail - easy. But we have a big problem."

Oliver sat down on a revolving leather chair, the steel wheels squeaking on the concrete floor. "I can't find the designs of their NextGen cars. No technical specifications, no engineering drawings n-o-t-h-i-n-g." "What does that mean?" Oliver seemed pretty calm about it all. "That means they are not on any systems that are connected to the Internet. They must be somewhere else. There's an air gap" he held his hands apart to illustrate a gap. Oliver clasped his hands behind his head. "Hold on… let me think."

He abruptly got up and straightened his shirt and hand-combed his thick brown hair. "Ok. No problem. Leave that to me." "What are you going to do?" Jordan was a tiny bit curious. "Don't worry about it. Oh – wait… do they know you are in their network?" "No, not yet. I will clean them out later. I will make it look like a big attack - then delete the logs so we disappear." "Perfect. Wait until I tell you." Oliver started walking to the door on the far side of the warehouse. He stopped abruptly and turned to Jordan. "Wait - so how did you get in? Did you use a zero-day vulnerability? Hijack a VPN connection?" he asked. Jordan smiled. "Nope - I saved you some money, mate. I pretended to be the anti-virus shop. I sent them an urgent virus update." "What was in the update?" "A real virus." Jordan cackled with laughter. Oliver strode out of the warehouse laughing to himself - humans truly are the weakest link.

## 9   Act 9

**Recent Past, Kuala Lumpur (Malaysia)**

"We can get you all the financials, but we have run into some unexpected developments regarding the engineering information," Oliver sat face-to-face with his Asian contact once again. Another cafe, another beautiful beach. It could have been anywhere. But the client insisted to meet in Kuala Lumpur. Maybe he felt it was far enough from his homeland but close enough that he could feel comfortable amongst the locals. Oliver didn't care. The client could have passed for a tourist from pretty much anywhere. "What do you mean? What unexpected developments?" the client was obviously not impressed given the amount of money he'd paid for Oliver's services. "The blueprints seem to be on a separate network that is not connected to the Internet." Oliver was hoping the client would not panic and walk away. It wouldn't have been the first time – in this business clients can be quite unpredictable. "So, I suggest that you simply ask Horizon for the engineering information," said Oliver. The client leaned back on his chair with a puzzled expression. "What do you mean?" "Offer their engineers a job. Hint that you are building something amazing and you need them to bring their 'A' game." "Is it wise for us to reveal ourselves to Horizon?" the client put on his sunglasses. "You already revealed yourself when you tried to take pictures of their factory floor. This is no different. And if they are like most busy firms oblivious to the world around them then they won't notice until they lose their first contract and it bites them in the ass." "I can't see Horizon's employees travelling halfway across the world to join us," the client seemed unconvinced. "I have built profiles on all the employees - there is one that is ripe for the taking. Make him an offer he can't refuse."

## 10 Act 10

**Present time, Hamburg (Germany)**

"Buzz", Anders pressed the doorbell of a townhouse softly in case there were young kids sleeping. He waited patiently at the front door, it was evening and very, very cold in Hamburg that day. The door eventually opened and with it came the welcome warmth of a cosy home. "Guten Abend... good evening Karim - it's been a very long time." 'Guten Tag - Anders! Habeebi - my friend, how are you?' Karim shook Anders' hand vigorously, a broad smile etched in his handsome face. "Its freezing outside, come in, come in," Anders was ushered into the living room where steaming mint tea and Turkish baklava awaited.

"I need your help Karim." Anders said once they were seated and sipping tea. "How is business? You're a highflying executive now. No one believed you'd actually build a simple passenger car, let alone a giant killer. Mabrook - congratulations." "Thank you - you believed in me when nobody would." Anders smiled. "What can I do for you?" Karim was always good at changing the subject. Anders took a deep breath and related the events of the last few days. Karim showed no emotion, he just listened intently, nodding his head along the way. He waited politely for Anders to finish. "Ok. Let me put on my security consulting hat, I need to ask you a few directed questions." "Of course, ask away." "What is the most valuable asset that sustains your competitive advantage?" "The cars of course. Each car is worth upwards of a million and a half US dollars." said Anders matter-of-factly. "But you are giving the cars away, so yes they are worth a lot but the cars themselves cannot be the asset, can they? As an automotive manufacturer, the *capability* to make cars that your market will buy at the price you set is your primary asset, yes?"

Anders thought for a moment. "Yes. It's our capability to make cars rather than the cars themselves that sustains our competitive advantage." "So... correct me if I'm wrong here, but Horizon cars are predominantly hand crafted, more so than say Lamborghini or Ferrari?" "Yes." Anders agreed. "Well then what is driving your competitive advantage is your specialized knowledge." Karim was stroking his salt-and-pepper beard with one hand and twirling a pencil with the other. "Knowledge? You mean information?" Anders looked a bit confused. "No, I do mean knowledge as opposed to information. Let me explain. Knowledge is much more than information. Its 'how' you make the car that matters, the 'how' points to more than just technical specifications, it's the experience and insight, the values and judgment which guides the actions and decisions of your engineers. The knowledge is in their heads, but it sounds like it's also embedded in the processes and in the specialized tools and techniques you've invented yourself to design, develop and manufacture successive generations of Horizon cars.

"So, I don't get it," Anders said. "What was the point in stealing the designs, the technical specifications, and the supplier and pricing information? That won't give them the capability anyway?" "Yes, but keep in mind FC Design is not starting from nothing. It sounds like they have most of the pieces of the puzzle including plenty of capability in automotive manufacturing, but they are missing the final piece in the puzzle - they still have to learn how to make cars like you do. The information they have taken is useful because it will help their engineers to understand in part how you build your cars. But they still have some way to go. The only way they would have been able to make the NextGen car before you is if they had at least some of your engineers on the factory floor to demonstrate Horizon practices."

And then it dawned on Anders. "So that's why he left!" he cried. "Who left?" Karim gave Anders a perplexed look. "Sorry - I forgot to relate that part of the story. The employee that left with our designs was an engineer from our manufacturing team." "That would do it," nodded Karim. "So, what do I do now?" "I'm not a business consultant, I'm a security consultant. I suggest you fix the leak — but look more broadly at people, process, and technology rather than just technology. It's the security of knowledge, not just information and its not limited to IT.Anders looked at Karim. "I don't think I can do this on my own."

## 11 Act 11

**Present time, Horizon Automotive's offices in Sweden**

Karim and Kurt sat around a large square table staring at a detailed network diagram of the Horizon network. "Well, what do you think?" asked Kurt. "First good thing is that R&D is designed to be completely segregated from the main network. But the rest of the organization has only one line of defence - the outside perimeter firewall. It seems you've adopted a castle defence strategy but with only one layer and a giant blindspot inside as you aren't monitoring internal activity at all. And by the way your firewall is a garden variety off-the-shelf product which almost certainly has bugs in it that outsiders know about. Other than that, you have the standard anti-virus software and standard access control with passwords.

"So, what to do?" Kurt was interested in solutions more than diagnosis. "Change the way you think about security before you change the infrastructure itself." "What do you mean?" Most organisations only have to worry about IT service availability so the threats they combat are accidents, natural disasters, and bots of various sorts. But you've got an intelligent and very resourceful adversary that is out to steal your capability rather than disrupt your IT services. And if they've done their homework then they are probably more knowledgeable about your network and systems as well as your personnel and your processes than you are. Karim paused to let this sink in, then continued.

For the majority of organisations facing a non-intelligent threat I'd just recommend shoring up the perimeter by penetration testing followed by heavy monitoring of incoming and outgoing traffic to maintain safe zones and to track control effectiveness. But in your case, I wouldn't recommend you setting up a castle to create a safe interior environment shielded from an unsafe exterior environment. Ultimately your plan was to sleep in peace with a sense of security around you, right? Kurt scratched his head and nodded. "How can they know more about our routines than we are?" said Anders. "They see things as they are, you see them as they should be," replied Karim philosophically. "So where do we start then?" "Everything starts with risk. Until you know what you are defending against, you can't develop an effective strategy."

## Copyright – Revised paper only

The following copyright paragraph must be appended to the paper after revision. Please ensure the hyperlink remains for electronic harvesting of copyright restrictions.

# Horizon Case Study: Teaching Note

Although technology-centered teaching is both necessary and critical to the development of future generations of IT security professionals, they are not particularly suited to management executives for two main reasons. Firstly, the IT security subjects focus exclusively on technology controls whereas security management professionals require a background in enterprise-level security management such as Strategy, Risk Management, Policy, and SETA (Security Education, Training and Awareness) in addition to the role of governance and culture (Maynard et al. 2011; Park et al. 2012). In particular, it is important that discussion of these aspects considers issues such as legal and regulatory directives, business impact and cost, stakeholder requirements including company directors, middle management and end-users (as well as outside stakeholders e.g. vendors and clients) and even the actions and intentions of competitors.

Secondly, IT security focuses on the security of IT assets in the digital environment such as networked systems and services and the information that flows through them, whilst largely ignoring the security of information in the physical environment where it is stored in hardcopy form and within the minds of personnel (Ahmad et al. 2005). The implications for business information security are considerable. Protecting the competitive advantage of organizations requires a comprehensive understanding of secure *knowledge* management, secure handling of hardcopy documentation and the ability to classify sensitive information- topics that are not typically covered in IT security subjects (Ahmad and Maynard 2014). Therefore, a broader enterprise view of information security is needed to protect Intellectual Property (IP) such as trade secrets and R&D data, an imperative for protecting innovation at an organizational and national level. To develop this area of expertise, an understanding of knowledge management is critical, especially the difference between tacit and explicit knowledge in addition to how knowledge is acquired, combined, and shared amongst human networks. Further, as Information Security breaches are often caused by employee failure to comply with procedures, it is particularly important for a subject in Information Security Management (ISM) to address the human dimensions of Information Security (Vance et al. 2012).

Drawing on the above discussion, teaching ISM should reflect the following principles:

P1. ISM protects the business function of the organization and therefore must consider strategy, risk, policy, training, governance, and culture
P2. ISM protects information and knowledge wherever it may be stored and however it may be transmitted
P3. ISM must address a range of threats both purposive and incidental, spanning technology, process, people, and information.
P4. ISM is essentially a people problem that has some technology solutions.

## 1 Design of the Horizon Case Study

Horizon is an automotive manufacturer that survives in a very competitive market due to its unique capability to consistently produce high-quality hyper cars. A competitor steals Horizon's trade secrets and subsequently develops and markets a rival product without Horizon noticing. The story features characters at senior and middle management level that engage in perfectly understandable and yet problematic behaviours from a security perspective. The case study highlights insecure behaviours that are typically found in organizations. Students with a professional background will have observed these or even engaged in such behaviours themselves, thereby creating empathy on the part of students for the central characters. The case study is written as a series of dialogues so that students hear the characters speak in their own voices – this creates a rapport between students and the characters allowing for an accurate understanding of the situation. The case also has pedagogical utility (see Table 1) and is applicable to organizations that have sensitive IP to protect against identified competitors.

The storyline was designed to complement the structure of an Information Systems Security curriculum. We used Whitman and Mattord (2017) as the principle textbook (see chapter mapping in Table 1 - pedagogical utility). The story consists of 14 acts and a set of discussion questions. Each act illustrates one or more key security management principles derived from research into management practice (see references in column 3 of Table 1). The acts have already been developed into an online clickable case study with accompanying high-quality graphics to enhance engagement. However only the text of the case study has been presented - due to copyright issues with the graphics.

| Act: Storyline | Pedagogical Utility |
|---|---|
| 1-3: Horizon learns it has lost a key contract to an unknown party that has stolen the design to its Next Gen hyper car. The CEO begins an investigation into how such a significant information security breach could have occurred. | Security Principles: P1, P3; Curriculum Topic: Introduction to Information Security (Ch 1, Whitman and Mattord 2017); Seminar Focus: Students reflect on the CEO's thought process and evidence of the breach as well as the role of the organisational security function and the fundamental distinction between data, information and knowledge assets (Shedden et al. 2016). |
| 4: Horizon's security investigation reveals the organization had been breached many times by FC Design, a new competitor in the hyper car market keen on developing its own competitive advantage. | Security Principles: P1; Curriculum Topic: Planning for Information Security (Ch 4, Whitman and Mattord 2017); Seminar Focus: At the conclusion of the investigation, students are asked to identify the deficiencies in Horizon's risk management processes (Webb et al. 2014). |
| 5-6: The story rewinds to a time in the past when FC Design planned its attack on Horizon. FC Design recruits a high-level intelligence operative with strong connections to the hacker community. | Security Principles: P3; Curriculum Topic: Need for Information Security (Ch 2, Whitman and Mattord 2017); Seminar Focus: Students gain an appreciation for the thought process of the attacking party as well as their strategy and tactics – these are typical of APT attacks (Ahmad et al. 2020; Ahmad et al. 2019). |
| 7: Horizon discovers that the responsibility to protect information assets is only partly addressed by IT and that the role and function of information security is the subject of widespread disagreement within the organisation. | Security Principles: P1; Curriculum Topic: Planning for Information Security (Ch 4, Whitman and Mattord 2017); Seminar Focus: Students reflect on how information security should function in an organisation and what capabilities and qualities should exist in a Chief Information Security Officer (Maynard et al. 2018). |
| 8-9: FC Design is only partially successful in procuring the ingredients needed to build their hyper car. They realise they have *know-what* but need a more creative approach to develop *know-how* or knowledge and subsequently their capability to manufacture hyper cars. | Security Principles: P2, P4; Curriculum Topic: None – this topic is not usually found in a college textbook; Seminar Focus: Students reflect on how to address the problem of knowledge leakage and to appreciate the security implications of the distinction between information and knowledge (Ahmad et al. 2014); Ahmad and Maynard (2014); Shedden et al. (2016). |
| 10: Horizon approaches a security consultant who asks insightful questions about how the organisation perceives the security challenge and what it does to mitigate risks. The CEO 'connects the dots' and finally sees the attack extent. | Security Principles: P1; Curriculum Topic: Security Risk Management (Ch 5, Whitman and Mattord 2017); Seminar Focus: Students reflect on limitations of adopting a narrow definition of Information Security especially where the primary asset is 'knowledge' (Shedden et al. 2016); (Park et al. 2012) |

**Table 1: Pedagogical Utility of the Horizon Case Study**

## 2   Utility of Fictional Storytelling in Executive Learning

We decided to develop our own fictional storytelling case drawing on real-world incidents (see legal indictments U.S. Department of Justice (2015),U.S. District Court for the Western District of Washington (2014), U.S. Department of Justice (2018)). There is in fact precedent to establish the use of entirely fictional cases. For example, the iPremier information security case published in the Harvard Business Cases specifically states "the iPremier cases do not describe real events and iPremier is not a real company" (Austin et al. 2002).

The Horizon case study features an interesting plot with elements of drama and suspense. The case study immerses students in a realistic scenario of a hyper car manufacturer that has suffered a loss of competitive advantage culminating out of a series of security incidents. The topic is interesting, provokes conflict, requires high-stakes decisions to be made, and is extremely relevant (the plot was drawn from news stories and relevant legal indictments). Management education has a long tradition of using case studies to impart knowledge. Typically, case studies are factual and do not incorporate fiction of any kind. However, Holtham (2015) has shown that fictional case studies that closely parallel real-world situations can deliver on pedagogical objectives as writers can use their creative license to craft a storyline that embeds particular principles, concepts and challenges they want to address. Most case study teaching is done face-to-face using cases written in text only. Video and graphics enhanced case studies make the problem more situated, which allows for greater realism, emotional complexity, and higher levels of engagement especially for a younger generation of learners.

The Horizon case study has several features that collectively meet good case study practices identified in recent research. The case study: (1) introduces students to the need for security management and high-level security management practices and (2) demonstrates the significance of theoretical principles discussed in the subject for example, the distinction between information and knowledge. For (1) the case has elements designed to assist students to engage in reasoning and theory-infused discussions that lead to improved ability to apply theory and practice principles. Evidence for this claim can be seen from the mapping of each act in the case study to specific papers in Table 1. Further, the case study elements are designed to bring authenticity to discussion and improve the level of engagement with the content. Individual sense making is also supported through built-in opportunities (e.g. dilemmas and insights) for instructors to facilitate reflection. For (2) the case study introduces the role of the 'security consultant' so that the student can learn from an example of mastery in security situations. The introduction of the security consultant allows the instructor to show how textbook principles and concepts are applied as 'best-practice' in a pseudo real-world situation.

Instructors can use Horizon as an aid for case-based learning in seminars and tutorials. To take best advantage of the case, we suggest instructors first map the specific Acts and discussion questions to the ISM topics in their curriculum. Once a topic has been delivered, instructors can ask students to read the relevant Acts after which the discussion questions can be used to create dialogue between theory and practice. This is particularly useful for executive students as Horizon can be used to instigate critical discussion and draw out relevant experiences. Over the course of the subject, students will find the Horizon storyline to be useful scaffolding upon which they can construct their own ISM learning.

## 3 Discussion Questions

| Act | Discussion Questions |
|---|---|
| 1 | 1. What do you think is going through Anders' mind at the end of act 1? |
| 2 | 1. What do you think is going through Anders' mind at the end of act 2? |
| 3 | 1. What is this meeting about? What can we deduce about Oliver's profession? Describe the firm represented by Oliver's client. What do we know about them and their strategic objectives? How are they planning to shift the competitive landscape? Why have they reached out to Oliver?<br>2. Discuss this category of threat. How would you describe this threat? How would you distinguish it from other types of threats? |
| 4 | 1. Why does Kurt think this is NOT an Information Security incident? What role does his team play with respect to security?<br>2. Which term – computer security, information security, information assurance best applies to this situation? Why?<br>3. What are the short-term and long-term costs of the leakage incident to the firm? |
| 5 | 1. FC Design had deliberately, systematically, and patiently targeted Horizon over a period of three to five years. Why was Horizon unable to mitigate the risk of leakage given each incident was actually reported? Relate your answer to the definition of 'Information Security' and ideas like 'risk perception'. |
| 6 | 1. What is Jordan's high-level game plan? What steps must he take to procure the targeted information? |
| 7 | 1. Suppose you were the Director of IT; how would you respond to Anders question?<br>2. Why doesn't Horizon have effective awareness training?<br>3. What qualities, capabilities, skills are required for a CSO - do you think appointing a CSO will solve the firm's problems? What will still remain?" |
| 8 | 1. How far up the kill-chain has Jordan progressed? What links remain to be traversed? |
| 9 | 1. What can organizations do to prevent employees from leaving and divulging information to a competitor? |
| 10 | 1. Explain the difference between Anders' and Karim's view of Horizon's primary asset. What is Karim's point? How does that change how Horizon develops its security program? |
| 11 | 1. Reflect on what further high-level strategy advice you would give Kurt. Think about the target of the adversary. What is the adversary after? Given the IT department is under-resourced, what is the simplest and most effective security advice you can give? |

Table 2: Horizon Case Study Discussion Questions